\documentclass[12pt]{iopart}
\usepackage{epsf}  
\begin{document}

\title{Hadrosynthesis at SPS and RHIC and the statistical model}

\author{Francesco Becattini\dag\  
\footnote[3]{e-mail: becattini@fi.infn.it}}

\address{\dag\ Universit\`a di Firenze and INFN Sezione di Firenze,
Via G. Sansone 1, I-50019, Sesto F.no, Florence, Italy}

\begin{abstract}
An analysis of hadron abundances in heavy ion collisions from 
SPS to RHIC energy within the statistical-thermal model is presented.
Pb--Pb collisions at 40 $A$ GeV are analysed for the first time here. 
Unlike stated in similar recent studies, the data analysis rules out 
a complete strange chemical equilibrium in full phase space. In fact,
the use of multiplicities integrated in full phase space or in a
limited rapidity window at SPS energy gives rise to different results
for the extra-strangeness suppression parameter $\gamma_S$ while
the extracted values of temperature and baryon-chemical potential 
do not vary significantly. This behaviour rises the question whether
the observed hadronic strangeness phase space saturation at RHIC 
within a small mid-rapidity window would hold in a possible full phase 
space analysis.  
\end{abstract}

\section{Introduction}

The statistical model of hadronisation, in different versions, is being used 
extensively in heavy ion collisions as a tool to determine the global conditions 
of the matter at the stage where interactions among hadrons cease (freeze-out), 
\cite{germ,cley,becah1,greece,gore,pbm2,becah2,pbm3,poland,rafe}. 
The agreement of the model with the data is somehow impressive \cite{cleysqm} 
and its validity has been unexpectedly proved in elementary collisions as well  
\cite{beca,beca2,becapt}, where inelastic rescattering between final hadrons is believed 
to be negligible. This latter finding has given rise to the interpretation that 
the hadronisation process itself generates hadrons according to the equiprobability 
of multi-hadronic phase space states (maximum entropy) \cite{beca2,stock,heinz} 
and the apparent chemical equilibrium among different species is driven by 
hadronisation itself rather than inelastic rescattering \cite{heinz}.    
   
The far most used observables in the statistical model analyses are hadronic
multiplicities and ratios of them. The main reason of this choice resides
in their Lorentz invariance, implying their independence on the momenta of
cluster (or fireballs) and thence on dynamical effects such as flow or collective 
momentum inherited by the pre-hadronisation dynamical evolution 
\cite{cleysqm,becapt}. 
On the other hand, both transverse and longitudinal momentum spectra do depend 
on dynamics besides the local properties of hadronising matter and the extraction 
of statistical-thermodynamical parameters based on them is more complicated. 
Moreover, they can only probe the state of hadronic matter at the later stage 
of {\em kinetic freeze-out} - i.e. when also elastic collisions cease - which 
does not necessarily coincide with the {\em chemical freeze-out} - i.e. when 
inelastic collisions cease - in heavy ion collisions.   

The detailed description of the statistical model which has been used for the
present analysis can be found elsewhere \cite{becapt,becah2}. As far as heavy ion
collisions, in the examined energy range, are concerned, the 4$\pi$ integrated 
multiplicities can be described by means of a grand-canonical formula depending 
on four free parameters (temperature $T$, baryon-chemical potential $\mu_B$, volume
$V$ and extra strangeness suppression parameter $\gamma_S$) provided that the
actual fluctuations of cluster masses and charges for fixed volumes meet 
a special requirement: they have to be the same relevant to the splitting of one
equivalent global cluster (EGC) into an arbitrary number of sub-clusters 
\cite{becapt}. If this is the case, the volume $V$ is meant to be the sum of the 
volumes of the clusters, measured in the EGC rest frame. 

The formulae and the methods used in this analysis are described in detail in 
ref.~\cite{becah2}. In this paper, we will mainly discuss the results obtained
in the analysis of three different systems: Pb--Pb collisions at $\sqrt s_{NN} = 
8.7$ and 17.2 GeV, and Au--Au collisions at $\sqrt s_{NN} = 130$ GeV.

\section{Analysis and results for SPS}

\begin{figure}
\begin{center}
\epsfxsize 9cm \epsfbox{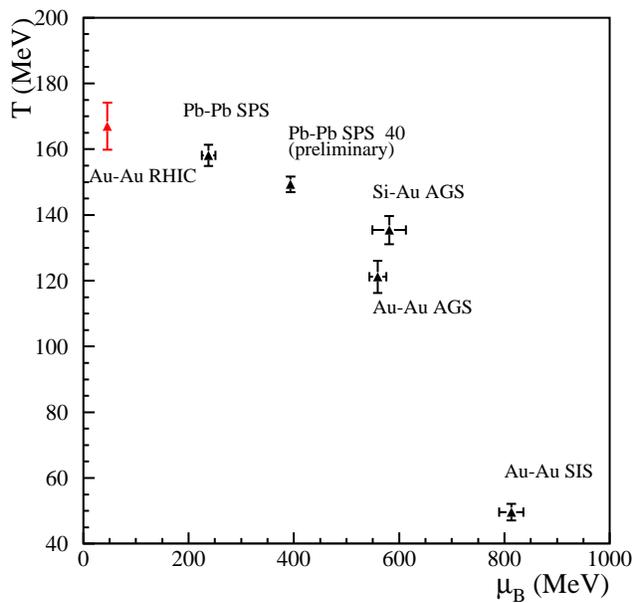}
\end{center}
\caption{\label{tmu} Temperatures and baryon-chemical potentials fitted 
with the statistical-thermal model analysis of hadron abundances in full 
phase space \cite{becah2}. For the RHIC point, hadronic multiplicity ratios 
at mid-rapidity have been used.}
\end{figure}

The basic analysis of Pb--Pb collisions at SPS has been performed with NA49
data measured at a beam momentum of 40 (preliminary) and 158 $A$ GeV corresponding to 
nucleon-nucleon centre-of-mass energies of 8.7 and 17.2 GeV. The multiplicities 
or their ratios
are measured or extrapolated to full phase space. The results are shown in
tables~1 and 2 along with relevant references. The fit quality is satisfactory in
both cases, taking into account that some measurements are still preliminary and
that the only sizeable discrepancy at 158 $A$ GeV, the $\bar\Lambda/\Lambda$ 
ratio \cite{becah2} is probably going to disappear \cite{privstock}. The $\gamma_S$ fitted 
values turns out to be in agreement with each other and less than 1 at both 
energies. This
clearly indicates that strangeness is not fully equilibrated in these collisions.
The value of temperature and baryon-chemical potential in Pb--Pb collisions at 
40 $A$ GeV, determined here for the first time, lie on the expected curve drawn by 
the values determined in previous analyses (see fig.~\ref{tmu}). It is worth
pointing out that recent measurements of short-lived strongly decaying particles
in NA49, such as $\Lambda(1520)$ \cite{friese,markert} have found a large discrepancy 
with respect to the prediction of the statistical model in this particular analysis
(1.45 $\pm$ 0.7 measured \cite{friese} against 3.516 expected). Within the 
statistical model, this can only be accounted for by admitting that the decay 
products of $\Lambda(1520)$ undergo significant elastic rescattering in the 
hadronic medium, thus losing memory of their invariant mass correlation.    

\begin{table}
\caption{\label{tab1}Fitted parameters in Pb--Pb collisions at a beam energy 
of 40 $A$ GeV and 158 $A$ GeV \cite{becah2} with full phase space multiplicities 
and in Au--Au collisions at $\sqrt s_{NN} = 130$ GeV with multiplicity ratios measured 
at mid-rapidity.}
\begin{indented}
\item[]\begin{tabular}{@{}l|llll}
\br
$\sqrt s_{NN}$ (GeV) & $T$ (MeV) & $\mu_B/T$ & $\gamma_S$ & $\chi^2/dof$ \\
\mr
8.7  & 149.3 $\pm$ 2.4 & 2.637 $\pm$ 0.068 & 0.822 $\pm$ 0.058 & 13.5/3 \\
17.2 & 158.1 $\pm$ 3.2 & 1.509 $\pm$ 0.075 & 0.789 $\pm$ 0.052 & 14.4/6 \\
\mr
130  & 167.0 $\pm$ 7.2 & 0.274 $\pm$ 0.030 & 1.04 $\pm$ 0.10 & 10.3/13 \\
\br
\end{tabular}
\end{indented}
\end{table}

The experiment WA97 \cite{wa97} has measured multiplicities of many strange 
particles in a limited phase space region ($\Delta y = 1$) around mid-rapidity 
in Pb--Pb collisions at 158 $A$ GeV in several centrality bins. In general, a cut 
on phase space like this, introduces a non-trivial dependence of the yields 
on the distribution of cluster charges (baryon number, electric charge and 
strangeness) as a function of rapidity, which can be otherwise disregarded in the 
$4\pi$ analysis \cite{becah1,becapt}. The statistical model is, by construction, 
not able to predict such distributions and, therefore, a dynamical model is 
needed. However, although an 
appropriate fit of abundances within the statistical model requires integration 
over $4\pi$, we have nonetheless performed a similar analysis especially in order to 
study the effect of the phase space cut upon the extracted parameters and compare 
our results with previous analyses where particle ratios measured in both full 
and limited phase space have been fitted at the same time \cite{pbm2}. 
\begin{table}
\caption{\label{tab2} Comparison between fitted and preliminary 
measured particle multiplicities in Pb--Pb collisions at a beam energy 
of 40 $A$ GeV. 
Results for 158 $A$ GeV can be found in ref.~\cite{becah2}.}
\begin{indented}
\item[]\begin{tabular}{@{}l|lll}
\br
 Particle         & Fitted  & Measured     & Reference \\
\mr
 $\pi^+$          & 264.6   & 282 $\pm$ 15   &  \cite{40gev1} \\
 $\pi^-$          & 293.3   & 312 $\pm$ 15   &  \cite{40gev1} \\
  K$^+$           &  51.91  & 56  $\pm$ 3    &  \cite{40gev1} \\
  K$^-$           &  19.53  & 17.8 $\pm$ 0.9 &  \cite{40gev2} \\
$\Lambda$         &  38.32  & 45.6 $\pm$3.4  &  \cite{40gev3} \\
 $\bar\Lambda$    &  0.7179 & 0.71 $\pm$0.07 &  \cite{40gev3} \\
 $N_p$            &  352.5  & 349 $\pm$ 5    &  \cite{40gev1} \\ 
\br
\end{tabular}
\end{indented}
\end{table}
\begin{figure}
\begin{center}
\epsfxsize 9cm \epsfbox{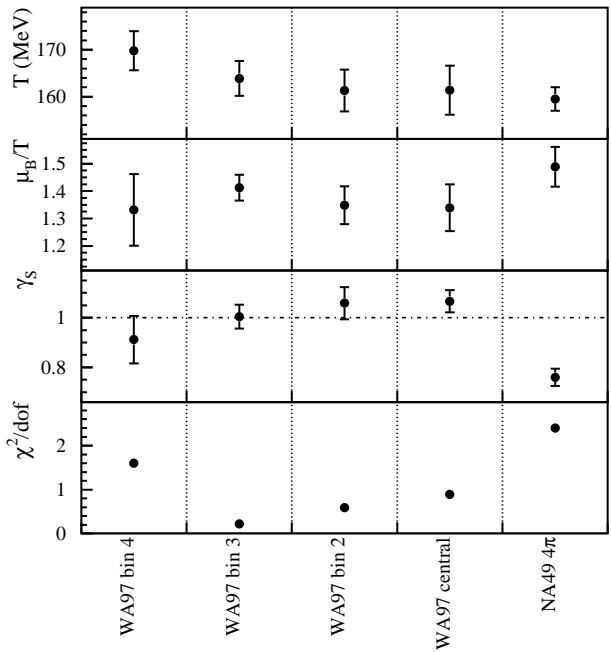}
\end{center}
\caption{\label{wa97} Statistical-thermal model parameters fitted by using
hadronic multiplicities measured by WA97 experiment in Pb--Pb collisions at
a beam energy of 158 $A$ GeV in various centrality bins (centered at 120,
204, 289 and 350 participant nucleons from left to right) compared with 
NA49 full phase space data fit.}
\end{figure}
It must be pointed out that, in order to keep the same number of free parameters of
the full phase space fit, this analysis has been performed by enforcing the 
net vanishing strangeness constraint, which might not apply in a limited kinematic 
region.

The results of the fits to WA97 multiplicities are shown in fig.~\ref{wa97} along 
with those relevant to the fit to $4\pi$ multiplicities measured by NA49 in the 
most central collisions. It can be clearly seen that the thermal-statistical 
parameters fitted by using
WA97 data do not essentially change going from central to peripheral collisions.
Moreover, the extra strangeness suppression parameter is about 1, unlike in the 
$4 \pi$ NA49 analysis. The different value of $\gamma_S$ fitted by using the limited 
and full phase space multiplicity samples proves quite definitely that ratios of 
hadronic yields measured in different kinematical regions cannot by any means be 
mixed in a single statistical model analysis, as has been done in ref.~\cite{pbm2}
where $\gamma_S = 1$ was claimed. It is also worth emphasizing that 
the fact that $\gamma_S <1$ in full phase space does not depend on the inclusion or 
not in the fit of particles carrying two strange quarks such as $\Xi$ and $\phi$ 
\cite{becah2}. 

\begin{figure}
\begin{center}
\epsfxsize 9cm \epsfbox{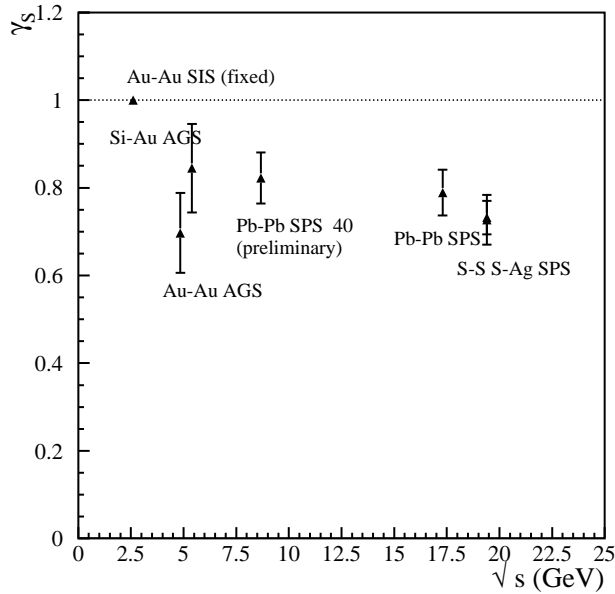}
\end{center}
\caption{\label{gs} Extra strangeness suppression parameter $\gamma_S$ fitted 
with the statistical-thermal model analysis of hadron abundances in full 
phase space \cite{becah1,becah2}.}
\end{figure}

\section{Analysis and results at RHIC}

We have used the same model to fit ratios of hadronic multiplicities measured
at RHIC in Au--Au collisions at $\sqrt s_{NN} = 130$ GeV in a small window 
around mid-rapidity ($\Delta y \simeq 1$), which is the probably the widest 
kinematical 
region accessible to RHIC experiments. As a limited region of phase space is 
involved, all the caveats discussed in the previous section concerning the use 
of such data hold. The data sample is the same as in ref.~\cite{pbm3} with one 
more measurement of $\phi/h^-$ ratio \cite{phi}. In this analysis, the net 
vanishing strangeness constraint has been used and hadronic weak decays products 
have been included in the multiplicities. The results are shown in table~1 and 
are in very good agreement with those in ref.~\cite{poland}. They are quite
consistent with ref.~\cite{pbm3} (taking into account that only 50\% of weak
decay products were included) while the temperature is somewhat lower than 
that quoted in ref.~\cite{kaneta}. 

\begin{figure}
\begin{center}
\epsfxsize 9cm \epsfbox{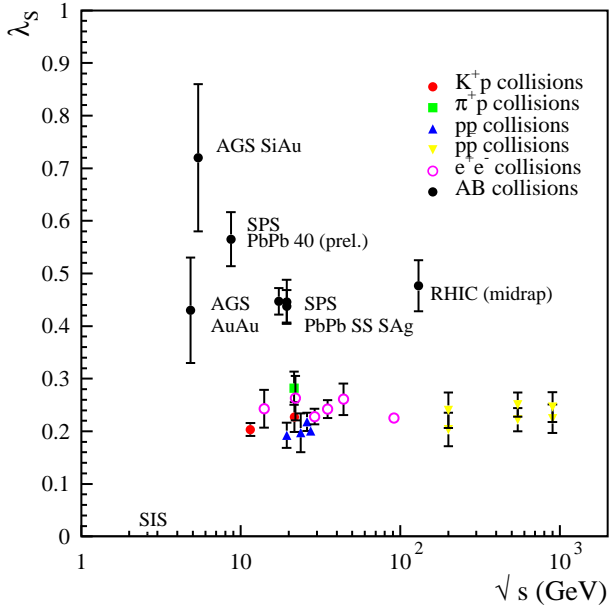}
\end{center}
\caption{\label{ls} Wroblewski factor $\lambda_S$ determined within 
the statistical model in several elementary \cite{beca2,becapt} and 
heavy ion collisions \cite{becah1,becah2} as a function of (nucleon-nucleon)
centre-of-mass energy. Unlike all other points, the RHIC value has been 
obtained by using mid-rapidity hadron yields.}
\end{figure}

\section{Discussion and conclusions} 

The obtained $\gamma_S$ values in several heavy ion collisions are shown
in fig.~\ref{gs} while the ratio between newly produced valence s quarks 
and u, d quarks for direct hadrons (the so-called Wroblewski factor) is
shown in fig.~\ref{ls} for both heavy ion and elementary collisions. This 
plot updates previous studies \cite{becah2}. It is clear that the behaviour
of strange quark production in heavy ion is very different from that in 
elementary collisions where $\lambda_S$ is seemingly constant $\simeq 0.2$.
The relative strangeness production seems to attain a maximum around a beam
energy of 30 $A$ GeV \cite{maximum} and decrease then to an asymptotic value 
of $\simeq 0.4$.
However, the largest energy point for heavy ion collisions in fig.~\ref{ls},
the RHIC one, has been calculated on the basis of a fit to mid-rapidity yields
and, thus, this $\lambda_S$ is not likely to be the same as in full-phase 
space on the basis of what has been discussed about Pb--Pb data at SPS in 
Sect.~2, rather un upper limit of it.  

\begin{figure}
\begin{center}
\epsfxsize 6cm \epsfbox{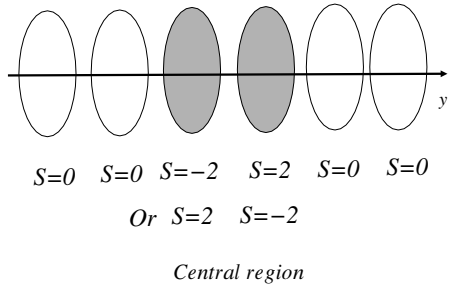}
\end{center}
\caption{\label{fireballs} The distribution of cluster charges as a function of 
rapidity may favour configurations with non-vanishing strangeness in the
central region, creating an apparent enhancement of strange particle production 
at mid-rapidity and faking $\gamma_S \simeq 1$.}
\end{figure}

A major issue in this kind of studies is related to the achievement or not of 
full hadron chemical equilibrium in heavy ion collisions. The analysis of 
$4 \pi$ multiplicities in heavy ion collisions clearly indicate that 
strangeness chemical equilibrium is {\it globally} not achieved in any of 
the examined systems (see fig.~\ref{gs}) \footnote{It has been proposed 
that even light flavours could be out of equilibrium inside the hadronising 
volume \cite{rafe} but, in our opinion, the present degree of accuracy in 
hadronic yield measurements does not allow to draw such a conclusion}. 
Recent claims that $\gamma_S=1$ in Pb--Pb collisions \cite{pbm2} depend
on a mixture of hadronic ratios measured in different rapidity windows.
The question now arises whether the full chemical equilibrium is at least achieved 
locally. The fact that $\gamma_S$ turns out to be about 1 by using hadronic 
yield ratios at mid-rapidity might be taken as a clearcut indication of the 
formation of a completely equilibrated hadron gas in the central region. 
However, it must be pointed out that this could not be the case.\\ 
The usual global statistical-thermal analysis of hadron multiplicities 
relies on a particular distribution of cluster charges and masses once their 
volumes are fixed \cite{beca2,becapt} but with no requirement on 
cluster charge distribution as a function of rapidity, provided that one deals
with full phase space multiplicities. Otherwise stated, the same $4 \pi$ yields 
can be obtained with many different distributions of cluster charges (e.g. 
strageness) as a function of rapidity and the ultimate reason of this special 
property is the Lorentz invariance of hadron multiplicities \cite{becah1,becapt}. 
It must be emphasized that this kind of requirement justifying the use of one 
global statistical-thermal formula for hadron abundances and based on charges 
and masses fluctuations of clusters for fixed volumes, is more general than 
the usual one based on the hydrodynamical picture. Indeed, in the latter argument, 
each cluster is described by grand-canonical parameters such as temperature 
and baryon-chemical potential and this implies that all of them have to be 
large enough; on the other hand, fluctuations of cluster masses and charges 
are anyway fixed by the grand-canonical framework, so the involved assumptions 
are altogether stronger than in the previous case. Furthermore, it might happen 
that clusters endowed with non-vanishing strangeness, or with a larger number of 
strange quarks to be coalesced in the final hadrons, tend to lie in the 
mid-rapidity region (see fig.~\ref{fireballs}), thus creating an apparent 
enhancement of strange particles thereabout even though the underlying $\gamma_S$
was less than 1 throughout. Therefore, one could obtain a $\gamma_S$ about 1 
by fitting mid-rapidity ratios even if this was not the case. Indeed, the dependence 
of fitted $\gamma_S$, as well as the other parameters, on the rapidity window 
has been studied in ref.~\cite{nuxu} on the basis of RQMD simulated data 
along with strangeness neutrality and a clear dependence of $\gamma_S$ on
the rapidity cut has been found.
 
In summary, the issues of local full chemical equilibrium and of the soundness
of the use of mid-rapidity multiplicities are not settled yet. More and more 
detailed studies are necessary, particularly in view of the experiments at 
RHIC and LHC where only a limited window in rapidity is accessible.   
  
\section*{Acknowledgements}

I am very grateful to the organizers of the conference for having
provided an enjoyable and stimulating environment. I am greatly indebted
with M. Ga\'zdzicki and R. Stock for having provided me with the NA49 
data and for their help in using it. Discussions with F. Antinori, 
J. Cleymans, A. Ker\"anen and N. Xu are gratefully acknowledged.   

\section*{References}

\end{document}